\documentstyle[prl,aps,multicol,epsf]{revtex}

\def\vc#1{{\bf #1}}
\newcommand{\be}{\begin{equation}}
\newcommand{\ee}{\end{equation}}
\newcommand{\bea}{\begin{eqnarray}}
\newcommand{\eea}{\end{eqnarray}}

\newcommand{\tjm}{$t$-$J$ }
\newcommand{\tttjm}{$t$-$t^\prime$-$t^{\prime\prime}$-$J$ }
\newcommand{\eps}{\varepsilon}
\begin{document}
\draft

\title{Stripes in Doped Antiferromagnets: Single-Particle Spectral 
Weight}

\author{Marc G.~Zacher, Robert Eder, Enrico Arrigoni,
Werner Hanke}
\address{Institute for Theoretical Physics, University of  W\"urzburg,
97074 W\"urzburg, Germany}

\maketitle

\begin{abstract}
Recent photoemission (ARPES) experiments on cuprate superconductors
provide important guidelines for a theory of electronic
excitations in the stripe phase.
Using a cluster perturbation theory, where short-distance
effects are accounted for by exact cluster diagonalization and
long-distance effects by perturbation (in the hopping), we
calculate the single-particle Green's function for a
striped \tjm model.
The data obtained quantitatively 
reproduce salient (ARPES-) features
and may serve to rule out "bond-centered" in favor of 
"site-centered" stripes.
\end{abstract}

\begin{multicols}{2}

There is by now
substantial experimental evidence \cite{tranquada,mook,ndlsco}
for a tendency of doped 
holes in high-temperature superconductors (HTSC)
to form stripes, leaving behind locally 
antiferromagnetic (AF) domains. 
The stripes can be 
static like in La$_{1.48}$Nd$_{0.4}$Sr$_{0.12}$CuO$_4$ 
(Nd-LSCO) \cite{tranquada} or dynamic like in 
YBa$_2$Cu$_3$O$_{7-x}$ \cite{sharma}. 
At present, there is an intensive
discussion and controversy,
whether stripes are directly connected with 
and beneficial for the 
microscopic mechanism in HTSC \cite{ichikawa,emery,zaanen,castellani,vojta,white2,wrobel}. 
What is clear, however, is 
that the apparent presence of static or dynamic stripes 
crucially influences low-energy excitations and thus
the foundations for such a microscopic 
theory. 
Evidence for this has recently been accumulated by 
angle-resolved photoemission spectroscopy (ARPES) both on 
the static stripes in the Nd-LSCO system \cite{ndlsco} and on 
dynamic domain walls in the LSCO compound \cite{ino,bianconi}. 
The electronic structure revealed by ARPES
contains characteristic features 
consistent with other cuprates, such as the flat band at 
low energy near the Brillouin zone face 
($\vc{k} = (\pi,0)$). 
In Nd-LSCO, the frequency-integrated spectral 
weight is confined inside 1D-segments in $\vc{k}$-space, deviating 
strongly from the more rounded Fermi surface expected from 
band calculations. 

\begin{figure}
\epsfxsize=8cm
\epsffile{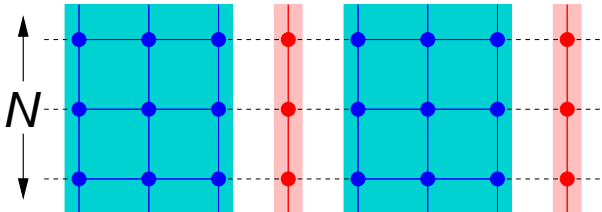}
\caption{Visualization of the cluster perturbation approach for stripes:
the groundstates for the half-filled
three-leg ladder ($3 \times N$) and the quarter-filled 1-leg chain 
($1 \times N$) are calculated exactly 
via exact diagonalization. The alternating clusters are then coupled 
via the inter-cluster hopping which is treated perturbatively.
}
\label{figcpt}
\end{figure}

In this Letter, we present
a numerical study of the 
single-electron excitations in a 
striped phase via cluster perturbation theory (CPT)
\cite{senechal} and a detailed comparison 
with recent ARPES results. The basic idea of our 
application of the CPT is indicated in 
Fig.(\ref{figcpt}): it is based on dividing the 2D plane 
into alternating clusters of metallic 
stripes and AF domains. 
The individial clusters are modeled by the microscopic
\tjm hamiltonian and solved {\it exactly} via exact diagonalization
(ED). Then the inter-cluster hopping linking the alternating
metallic and AF domains is incorporated {\it perturbatively}
via CPT 
on the basis of the {\it exact}
cluster Green's functions thus yielding the spectral function
of the infinite 2D plane in 
a striped phase.
Using this CPT approach, the important short-distance interaction
effects within the stripes are taken into account exactly while
longer-ranged hopping effects are treated perturbatively.
A study of the properties of experimentally observed 
stripe phases {\it solely} by ED\cite{maekawa}
is precluded by the prohibitively large unit cells.
The manageable clusters
for ED are simply too small to accommodate even a single such
unit cell.

\begin{figure}
\epsfxsize=8cm
\epsffile{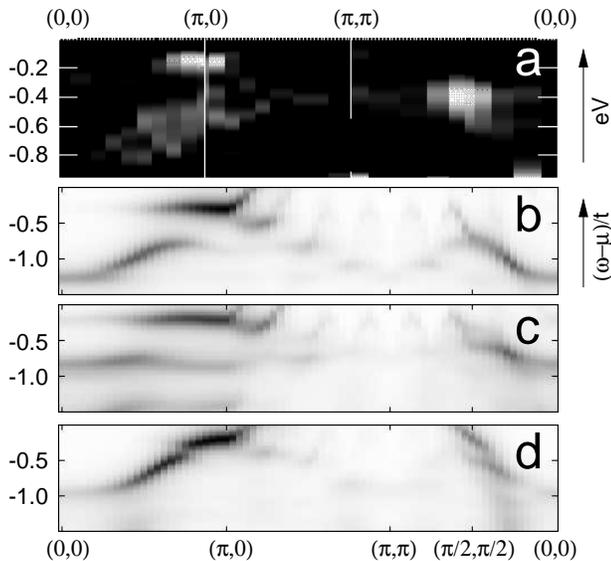}
\caption{ARPES results for La$_{2-x}$Sr$_x$CuO$_4$, theoretical single particle
spectral function $A(\vc{k}, \omega)$: 
(a) displays experimental ARPES results by Ino {\it et al.}
{\protect\cite{ino}}. 
The grayscale corresponds to the second derivative of the original
measured data. Flat regions are black, regions with high curvature (i.e. peaks) 
are white.
(b),(c),(d) show the results of the stripe CPT calculation for
$A(\vc{k}, \omega)$ 
(b: "$3 + 1$" site-centered;$t^\prime = 0$, c: "$3 + 1$";$t^\prime = -0.2 t$,
d: "$2 + 2$" bond-centered;$t^\prime=0$).
Here, $A(\vc{k}, \omega)$
is plotted directly with maximum intensity corresponding to black.
}
\label{figcompino}
\end{figure}

\begin{figure}
\epsfxsize=8cm
\epsffile{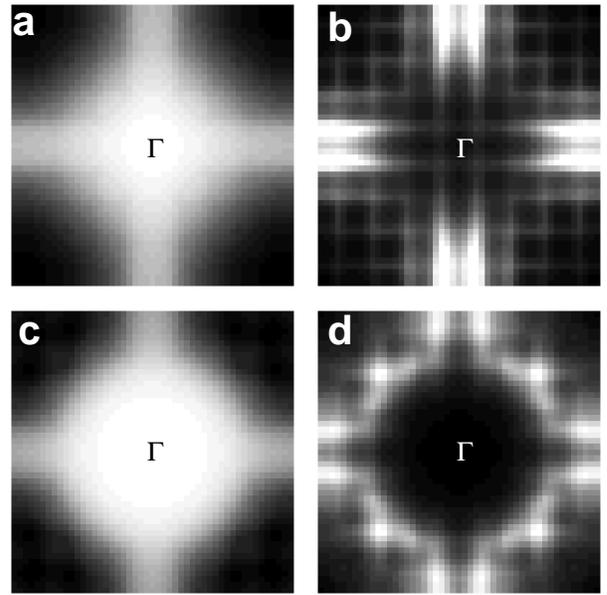}
\caption{Integrated spectral weight of site-centered (a,b) and bond-centered (c,d)
stripe configurations;
(a,c) total integrated weight in photoemission ($n(\vc{k})$), 
(b,d) low energy excitations 
(integrated weight in $E_F-0.2 t < \omega < E_F$).
The data are plotted for the whole Brillouin zone with the $\Gamma$-point in
the center. The result of the stripe calculations have been symmetrized to account 
for the differently oriented stripe domains in real materials. Regions of high spectral
weight correspond to white areas. 
}
\label{figfermi}
\end{figure}

Our main results are: 
(i) close 
to $\vc{k}= (\pi,0)$ we see, like in 
experiments, a two-component electronic feature (see Fig.(\ref{figcompino})): 
a sharp low-energy feature close to $E_F$ and a more broad 
feature at higher binding energies. Both features can be 
explained by
the 
mixing 
of metallic and antiferromagnetic bands at this $\vc{k}$-point.
(ii) 
the excitation near $(\pi/2,\pi/2)$
is at higher binding energies than the low-energy excitation
at $(\pi,0)$ and of reduced weight.
(iii) the integrated 
spectral weight of the cluster-stripe calculation 
resembles the quasi-one-dimensional segments in 
momentum space (see Fig.(\ref{figfermi}a)) as seen in the
Nd-LSCO experiment. 
Also in agreement with the Nd-LSCO experiment our calculation
finds the low-energy excitations 
near $(\pm \pi,0)$ and $(0,\pm \pi)$ (Fig.(\ref{figfermi}b)).
Interestingly, this agreement with 
experiment occurs only for so called "site-centered" 
metallic stripes (as shown in Fig.(\ref{figcpt})) and not 
for "bond-centered" metallic stripes. 
This seems important since it has been argued \cite{lee}, that,
for bond-centered stripes, 
superconductivity is expected to survive stripe ordering.
In the DMRG calculations by White and 
Scalapino \cite{white1} as well as in
dynamical mean-field (DMFT) studies
by Fleck {\it et al.} \cite{mfleck},
bond-centered and site-centered 
domains are very close in energy (ground-state). However, 
our technique allows to distinguish them dynamically. 

The computational technique for our calculation of the 
single-particle spectral weight $A(\vc{k}, \omega)$ and the 
Green's function is illustrated in Fig.(\ref{figcpt}). 
We solve the 
AF cluster (here, $N \times 3$) and the metallic 
cluster (here $N \times 1$ with a hole filling $n_h 
= 0.5)$ by ED and combine the individual clusters to 
an infinite lattice via CPT
as in the homogeneous case. Where it was technically feasible,
we extended the unit cell by a factor of two and diagonalized
two AF 3-leg clusters with a staggered magnetic field pointing in
opposite directions, resulting in a $\pi$-phase shifted
N\'eel order in the final configuration.
This site-centered "$3 + 1$" configuration with  
$\pi$-phase shifted N\'eel order 
of stripes was first suggested
by Tranquada {\it et al.} \cite{tranquada}.
Bond-centered stripes, on the other hand, are modeled by
2-leg ladders with alternating filling (half-filled, $n_h=0$ 
and doped, $n_h=1/4$). In the following we will refer to this
bond-centered configuration as "$2 + 2$".

Holes can propagate out of the metallic stripes into 
the AF insulating domains via the inter-cluster hoppings. 
In the Hubbard model, 
these hoppings correspond to one-body operators and can be 
treated within a systematic strong-coupling perturbation 
expansion. For homogeneous systems such an expansion was 
constructed in ref.\cite{senechal}. We 
take the lowest order contribution: 
\begin{equation}
G_\infty (\vc{P},z) = \frac{G_{cluster}(z)}{1 - 
\eps(\vc{P}) G_{cluster}(z)},
\label{eqcpa}
\end{equation}
which is of RPA form.
Here, $\vc{P}$ is a superlattice wave vector and $G_\infty$ 
is the Green's function of the "$\infty$-size" 2D system, 
however, still in a hybrid representation: real space 
within a cluster and Fourier-space between the clusters. 
This is related to the fact that $G_\infty(\vc{P}, z)$ is now an 
$M\times M$ matrix in the space of site indices (in the 
inhomogeneous stripe configuration of Fig.(\ref{figcpt}) $M 
= N \times 3 + N \times 1 = 4 N$). Likewise, $\eps(\vc{P}
)$ and $G_{cluster}$ are $M\times M$ matrices in real 
space with $\eps(\vc{P})$ standing for the perturbation, 
which includes hoppings out of the clusters. A true Fourier 
representation of $G_\infty$ in terms of the original 
reciprocal lattice then yields the lowest order CPT 
approximation. As discussed in ref.\cite{senechal} for the 
homogeneous case, the approximation in eq.(\ref{eqcpa}) is exact 
for vanishing interaction.
When the interactions are turned on, 
eq.(\ref{eqcpa}) is no longer exact (apart from the local limit, 
i.e. $t=t^\prime=0$), but strong interactions are known to 
be important mainly for
short-range correlations. These correlations are 
incorporated with good accuracy in modest-size clusters and 
are treated here by ED within the cluster. 
It has been shown by S\'en\'echal {\it et al.} 
\cite{senechal} that the CPT reproduces the spectral weight 
of the 1D and 2D Hubbard models in quantitative agreement 
with exact results.

To allow for larger cluster sizes $N$, we diagonalize 
the \tjm model 
and take its spectral (Green's) function as 
our local $G_{cluster}$ in eq.(\ref{eqcpa}) as an approximation
to the Hubbard model's (cluster-) Green's function. A comparison
of the Hubbard and \tjm model's spectral function on small clusters 
\cite{eskes} shows
that the strong low energy peaks have similar dispersion and weight
in both models,
the main difference being a transfer
of incoherent high energy spectral weight from momenta near $(\pi,\pi)$ to $(0,0)$.
The \tjm hamiltonian is defined as
\be
H = - \sum_{ij,\sigma} t_{i,j} \hat c^\dagger_{i,\sigma} 
\hat c^{}_{j,\sigma} + J 
\sum_{<i,j>} ( \vc{S}_i \vc{S}_j - \frac{n_i n_j}{4} ).
\ee
The hopping matrix element $t_{i,j}$ 
is nonzero only for nearest ($t$) 
and next-nearest neighbors ($t^\prime$).
The second sum counting the Heisenberg interaction $J$ runs over
all nearest neighbor pairs.
No 
double occupancy is allowed. 
We have chosen commonly accepted values for the 
ratio $t^\prime/t = -0.2$, $J/t = 0.4$, where 
$t \approx 0.5 eV$. In this Letter we present calculations
for systems with $N=8$ based on diagonalizations of a 
$N \times 3= 24$-site half-filled 3-leg ladder and a quarter-filled $8$-site
chain. Results for smaller $N=6$ do not differ much from $N=8$ results.
$N=6$, however, is somewhat pathological,
 since it has an odd number of electrons
in the quarter-filled chain.

We proceed to the discussion of the spectra:
Fig.(\ref{figcompino}a) shows the experimental ARPES results 
for LSCO at the superconductor-insulator transition
(doping $x = 0.05$) \cite{ino}. 
To enhance the structure in the obtained spectra, the authors of
ref.\cite{ino} plotted the second derivative of the ARPES spectrum, so 
areas with high second derivative are marked white and areas with low curvature
(i.e. flat intensity) are black.
This result is compared with the theoretical CPT calculation
for different stripe configurations with overall doping of $x=1/8$.
Figs.(\ref{figcompino}b,c)
 are for the "$3 + 1$" site-centered configuration.
Fig.(\ref{figcompino}b) 
shows the result for a "$3 + 1$" configuration
with alternating (i.e. $\pi$-phase shifted)
N\'eel order between the
3-leg ladders (induced by a staggered magnetic field
$B = 0.1 t$)
without next-nearest neighbor
hopping, Fig.(\ref{figcompino}c) shows the result for the "$3 + 1$" stripe
configuration with next nearest-neighbor hopping $t^\prime = -0.2 t$, however
without N\'eel order (due to the reduced symmetry, a $t^\prime$ diagonalization
with staggered field is not technically feasible). 
Fig.(\ref{figcompino}d) shows the result for a bond-centered "$2 + 2$" stripe
configuration.
We observe that the spectra 
for the site-centered
"$3 + 1$" configuration (Fig.(\ref{figcompino}b,c))
are in surprisingly good agreement with 
experiment, 
and, that the $t^\prime = 0$ calculation (Fig.(\ref{figcompino}b))
results in much more coherent bands due to the 
enforced N\'eel order in the 3-leg
ladders. 
Similar to recent DMFT calculations \cite{mfleck},
the sharp excitation near the Fermi surface around $(\pi,0)$, 
that has been interpreted by Ino {\it et al.} as the 
quasiparticle peak in the SC state,
is visible  as well as 
a dispersive band at higher binding energies which
(at least away from $(\pi, 0)$; see discussion below) can be
interpreted as remnants of the insulating valence band resulting from the
AF domains.
Especially in the N\'eel ordered configuration,
we observe a very coherent 
and pronounced band.
This clear dispersion is also
visible in the $(\pi,\pi)$ direction near $(\pi/2, \pi/2)$.
We note that, in agreement with the experimental result (Fig.(\ref{figcompino}a)),
the excitation at $(\pi,0)$ is at significantly lower binding energy
than the excitation at $(\pi/2,\pi/2)$. Neither the 2D \tjm model nor a
2D \tttjm model, with its parameters fitted to the insulating
state, can reproduce this result \cite{ttj}. This is a crucial effect of
the stripe assumption: With the stripes oriented along the $y$-direction,
the metallic band is dispersionless in $x$-direction.
Therefore, at $\vc{k}=
(\pi,0)$, the
minimum of the metallic spinon
band (located at $(k_x, 0)$ for {\it any} $k_x$) hybridizes with the top of
the insulating valence band resulting in a two-peak structure
with one peak
pushed to higher and the other pushed to lower binding energies
(see below).
At $\vc{k}= (\pi/2, \pi/2)$, on the other hand, the metallic band has crossed the
Fermi surface (its $k_F$ being $\pi/4$) and no mixing takes place.
Finally, the "$2 + 2$" bond-centered stripe configuration (Fig.(\ref{figcompino}d))
does not show much resemblance to the experimental result. Its main 
band is
much more two-dimensional, normal metal-like, 
comparable to the dispersion of a 2D tight-binding band.

Fig.(\ref{figfermi}a) plots the integrated spectral 
weight 
$n(\vc{k})$
for the "$3 + 1$" 
site-centered stripe configuration with $t^\prime=0$. 
Although not as clear as in the Nd-LSCO ARPES experiment
(from ref.\cite{ndlsco}),
the "Fermi surface" is rather one-dimensional in structure.
Like in Nd-LSCO the low energy excitations (shown in Fig.(\ref{figfermi}b),
calculated by integrating over a $\Delta \omega = 0.2 t$ window below
the Fermi-energy for each $\vc{k}$-point in the Brillouin zone)
are located 
near the $(\pm \pi,0)$, $(0,\pm \pi)$ points in momentum space.
In Fig.(\ref{figfermi}b) we notice the $8 \times 8$ square lattice
of bright points. This is the repeated Brillouin zone of the supercell
consisting of $(3 + 1 + 3 + 1) \times N = 8 \times 8$ 
lattice sites (due to the N\'eel order in $x$-direction). Clearly, 
the low energy excitations in the momentum space of the supercell
are near the $(\pm \pi,0)$ and $(0,\pm \pi)$ points as well.
Fig.(\ref{figfermi}c) 
shows the integrated spectral weight $n(\vc{k})$ of the "$2 
+ 2$" \tjm stripe calculation. Here, the 
CPT "Fermi surface" is much more rounded, similar to the 
quasi-2D Fermi surface known from band calculations. 
The low energy excitations are located isotropically around 
the "Fermi surface" as well (Fig.(\ref{figfermi}d)).
The loss of one-dimensionality observed for 
this bond-centered "$2 + 2$" stripe configuration is accompanied by
a substantial enhancement of spectral weight
near $(\pi/2, \pi/2)$. From these 
$A(\vc{k}, \omega)$ results we conclude that, at least in 
the Nd-LSCO system, the stripes are site-centered and of 
"$3 + 1$" type.

\begin{figure}
\epsfxsize=7.5cm
\epsffile{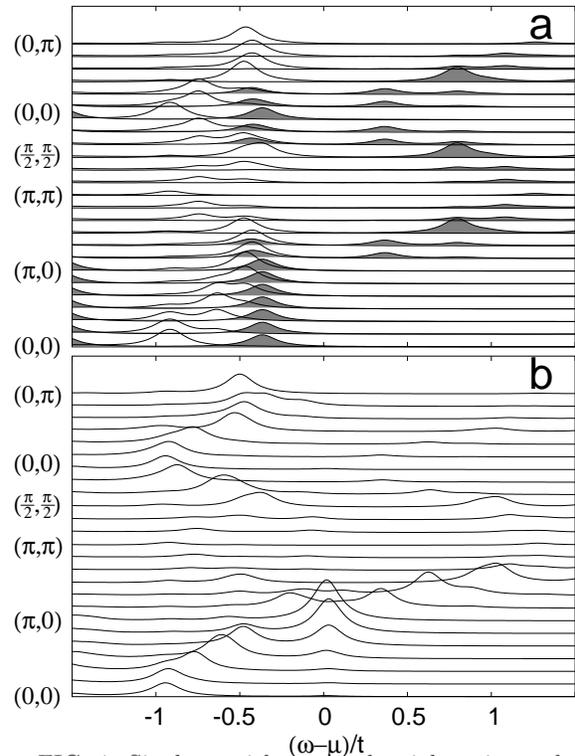}
\caption{Single particle spectral weight prior and after application of the
cluster perturbation theory:
In (a) the shaded curve gives the spectrum of the 1D metallic chains
and the solid line corresponds to the spectrum of the uncoupled AF domains
(see Fig.{\protect \ref{figcpt}}).
In (b) the result of the cluster perturbation is plotted.
}
\label{figline}
\end{figure}

With our technique we are able to resolve for each excitation, 
whether its
main origin is from the insulating or the metallic part of the stripe configuration:
In Fig.(\ref{figline}) we compare the spectra for the unperturbed stripe
configuration with inter-cluster hopping set
equal to zero (in 
Fig.(\ref{figline}a): solid curve for the AF domains, shaded curve for the 1D metal)
with the result of our CPT calculation (solid line in Fig.(\ref{figline}b)) 
with inter-cluster hopping 
$t$ (for the N\'eel ordered "$3 + 1$"
configuration). All of the spectral weight
 in inverse photoemission ($\omega > 0$)
naturally stems from the chain since the half-filled 3-leg \tjm ladder does not
have target states for an inverse photoemission process.
The stripes are oriented along the $y$-direction. Therefore, we can conclude that
peaks (in the calculation with $t=0$), 
that show a dispersion along $(0,0)$ to $(\pi,0)$ direction
stem from the 
AF domains (solid line in (\ref{figline}a)) whereas the metallic excitations
prior to the mixing (shaded curve in (\ref{figline}a))
are dispersionless.
By comparing the three curves, we 
therefore  conclude that the sharp
quasiparticle peak near $(\pi,0)$ results from the mixing of the (dispersionless)
$(k_x,0)$ minimum of the metallic spinon
band and the top of the insulating valence band
situated at $(\pi,0)$. Going from $(\pi,0)$ to $(\pi,\pi)$, the metallic band
becomes dispersive and crosses, in agreement with
experiment \cite{ndlsco,ino}
 the Fermi surface at $k_y = \pi /4$ (since it is
quarter-filled). The dispersion of the insulating band, however, is in the opposite
direction. For this reason, in the final spectrum, we observe that the sharp
quasiparticle peak at $(\pi,0)$ becomes dispersive going into $(0,\pi)$ direction
 and eventually crosses the Fermi surface, however, with diminishing weight due to
the absence of mixing with the insulating band.
This effect is best visible in the grayscale plot of Fig.(\ref{figcompino}b).
This finding
may serve to clarify questions raised in the experimental ref.\cite{ndlsco}
concerning the origin of the quasiparticle peak at $(\pi,0)$.

To summarize, the single-particle spectral weight 
$A(\vc{k}, \omega)$ 
was calculated 
for different stripe configurations
employing an application of the 
cluster perturbation technique for inhomogeneous systems.
This technique allows to 
obtain all $\vc{k}$-points and, therefore, allows for a detailed 
comparison with ARPES data.
The $A(\vc{k}, \omega)$ results for 
the "$3 + 1$" site-centered configuration display 
salient features observed in experiments such as a two-peak 
structure around $(\pi, 0)$ with a sharp excitation
close to the Fermi-energy and a broader feature at higher binding energies,
a quasi-1D distribution of spectral weight, 
and low energy excitations located
around the $(\pi,0)$-points in the Brillouin zone.
The theoretical results suggest to rule out 
the alternative bond-centered "$2 + 2$" configuration.

The authors acknowledge 
financial support from
BMBF (05SB8WWA1) and DFG (HA1537/17-1). The calculations 
were carried out at the high-performance 
computing centers HLRS (Stuttgart) and LRZ (M\"unchen).

\narrowtext

\end{multicols}

\end{document}